\documentclass[twocolumn, journal]{IEEEtran} 
\makeatletter 
\def\ps@headings{%
\def\@oddhead{\mbox{}\scriptsize\rightmark \hfil \thepage}%
\def\@evenhead{\scriptsize\thepage \hfil \leftmark\mbox{}}%
\def\@oddfoot{}%
\def\@evenfoot{}} 
\makeatother 
\usepackage{latexsym,times,epsf,amsmath,amssymb,amsfonts,graphicx}
\usepackage{epstopdf}
\usepackage{graphicx}
\usepackage{subfigure}
\usepackage{multirow}
\usepackage{cite}
\usepackage{setspace}
\usepackage{comment}
\usepackage{float}

\usepackage{algpseudocode}
\usepackage{algorithm}
\usepackage{mathrsfs}
\usepackage{array}
\renewcommand{\baselinestretch}{0.94}

\begin{document}
\title{ECG Signal Compression and Optimization in Remote Monitoring Networks}
\author{Pengda Wong}
\maketitle

\begin{abstract}
We proposed a practical ECG compression system which is beneficial for tele-monitoring cardiovascular diseases. There are two steps in the compression framework. First, we partition ECG signal into segments according to R- to R-wave periods. The partition aims at achieving more stable statistical features between segments of ECG signal which is beneficial for saving bit rates. After the partition, we optimize the bit rate in the sense of minimizing ECG reconstruction error under a constraint of consumed bits. From the experiment results, the proposed compression scheme is able to reduce the computation for updating codebook, and save channel capacity resources for transmitting ECG signals. 
\end{abstract}

\section{Introduction}
\label{sec:introduction}

Cardiovascular disease~(CVD) is the leading cause of death in the United States which is also a major threaten to health in other areas of the world. Cardiovascular telemonitoring is an effective and economic approach to providing instant CVD diagnosis and alert. The instant operation is important to reduce the potential harm caused by CVD. Furthermore, ECG signal is an effective tool of helping us diagnose cardiovascular diseases. Therefore, remotely recording and transmitting ECG signal have significant importance for providing the instant operation in CVD diagnosis and alert. With the fast development in signal processing and communication technologies~\cite{Nevat2016,Szurley2017,Huang2014, Huang2011}, the remote monitoring is near to practice. However, instant remotely monitoring ECG signal caused cost is a major challenge in the practice. The improvement of bandwidth efficiency is an effective approach to reducing the cost. For this purpose, we study the ECG compression scheme in this paper.



Standalone compression of ECG signal is widely studied. In~\cite{Marisa2015}, Hermition polynomials are employed to compress ECG signal. Four Hermitian polynomials are driven to constitute a set of orthogonal basis which spans a space in four dimensions. ECG signal is projected onto four-dimension space for compression. Indeed, ECG waveforms have large diversity for different people or different health conditions of a same person. Thus, the four-dimension space can hardly provide a compression with high fidelities. 

As a widely utilized signal processing tool, compressed sensing is also leveraged to compress ECG signal~\cite{Ravelomanantsoa2015,Ravelomanantsoa2014}. In~\cite{Ravelomanantsoa2015}, a deterministic matrix was proposed to map ECG signal to a space in which ECG signals has sparse expression. The similar strategy is also taken in~\cite{Ravelomanantsoa2014}. Indeed, one deterministic matrix can hardly guarantee the projected ECG signal of different persons has a sparse form. In~\cite{Polania2015}, wavelet transform is employed to build the mapping matrix for ECG compressed sensing. Different from the work in~\cite{Ravelomanantsoa2015,Ravelomanantsoa2014}, not the original ECG signal, but difference between consecutive two ECG segments are compressed in~\cite{Polania2015}. The segmentwise differential ECG compression is also adopted in~\cite{Mamaghanian2011}. However, ECG signal is assumed to be periodic~\cite{Polania2015}. As we know, ECG signal is far from being stationary or periodic. In~\cite{Zhang2015}, compressed sensing built on $\ell$-1 norm was utilized to compress ECG signal which focuses on reducing the coherence between the measurement matrix and reconstructed signal. The presumption of ECG being stationary is also taken in~\cite{Zhang2015} which is far from being real case.  

In our analysis, we remove the assumptions that ECG is stationary or periodical. Afterwards, we optimize the frame-level compression of ECG signal. In our compression scheme, R-waves of ECG signal are first detected. According to the detected R-to-R periods, we divided ECG signal into segments which have unequal length in usual cases. Afterwards, the ECG segments are converted into the ones with equal length via interpolation or decimation. The converted ECG segments are packaged into frames. Next, we optimize the bit rate in frame-level ECG compression at the purpose of minimizing reconstruction error with constraint of bit number. More explicitly, the contributions of this paper are listed as follow,

\begin{enumerate}
  \item We investigate the optimum size of ECG segments. We investigate two key issues affecting the selection of segment size. First, under the presumption of equal length partition, the longer ECG segments have more stable inter-segment statistical features. On the other side, computation burden exponentially increases with the extension of segment length. Second, R- to R-period estimation assisted ECG partition is able to achieve the stable inter-segment statistical feature at shorter average segment length. The short length means less computation.

  \item We propose a novel R-wave detector which is insensitive to the diversity of ECG waveform. Generally, shapes of ECG signal are different from person to person. Even for a same person, signals from different ECG leads in a standard ECG system are not in the same shape. Our proposed R-wave detector manages to convert ECG signals with diverse shapes into waveforms in a closely similar form. This conversion enables the detector stronger feasibility for ECG signal in diverse shapes. 

  \item A segmentwise ECG compression scheme is proposed which optimizes bit rate. The proposed compression scheme provides us a practical approach to reducing compression error under a constraint of bit rate. 

  \item We design three structures based on which our bit rate optimization scheme can be implemented. 

\end{enumerate}


In Section~\ref{sec: related}, the work related to QRS wave detection and bit rate allocation for ECG transmit is introduced. After analyzing the relationship between ECG distribution per segment and segment length, we propose a novel ECG partition method which relies on R-to-R period estimation in Section~\ref{sec: seg_len}. To improve the accuracy of the period estimation, a novel R-wave detection method is proposed in Section~\ref{sec: QRS detect} and this method is insensitive to the change of ECG waveforms. After the R-to-R period based partition, a practical compression method is proposed in Section~\ref{sec: bit rate allocation}. Furthermore, for a more intuitive impression on our proposed ECG compression scheme, block diagrams of the compression scheme in three types are presented in Section~\ref{sec: ecg compression}. Afterwards, simulations are performed to illustrate the effectiveness of our work along with the results in Section~\ref{sec: experiment} which is followed by conclusions in Section~\ref{sec:conclusion}.


\section{Related Work}
\label{sec: related}





In literature, there are a large number of reports on R-wave detection. A level crossing rate based analog-to-digital converter~(ADC) is proposed to detect the R-wave in~\cite{Ravanshad2014}. Indeed, the level crossing rate changes at different ECG signals. Thus, uniform quantization levels in the level crossing rate based ADC are difficult to determine. The selection of quantization levels has not been introduced in~\cite{Ravanshad2014}. QRS complexes are detected in~\cite{Deepu2015} via counting the numbers of continuous ECG rising and falling in ECG waveforms. The detection consists of two sequential steps. In the first step, the number of continuous ECG rising events is counted. Once the number hits a threshold, the detection procedure goes into the second step that the number of continuous ECG falling is counted. Only if both the number for rising and that for falling reach their thresholds respectively, a QRS complex is claimed to successfully detected. Indeed, a general threshold is difficult to be selected due to the significant change in different ECG signals. In~\cite{Wei2014}, iterative FFT is applied in the heart rate estimation which determines the heartbeat period by detecting the spectrum peak of ECG signal. The spectrum peak detection based method could be available in estimating the average of R-to-R periods, but not a good choice for the estimation of separate R-to-R periods. Heartbeat rate was detected based on rule of maximum a posteriori~(MAP) in~\cite{Sprager2014}. The MAP based estimation relies on a distribution model of ECG samples. Due the same reason of the diversity in ECG signal, the distribution of ECG can hardly be represented by one general model. 






As for the ECG compression methods, there is one categorization way, fixed and varying bit rate. Fixed bit rate for ECG compression can be implemented using a simple structure since a constant number of bits are used in each frame. The discussions on ECG compression with fixed bit rate are performed in a relatively larger number of papers, such as~\cite{Ravelomanantsoa2015,Marisa2015,Deepu2015}. On the contrary, reports on the allocation of varying bit rate in ECG compression is limited. The work in~\cite{Kim2006} discusses the selection of segment size which does not analyze the allocation of bit rate. Varying bit rate framework is taken as the research background in~\cite{Kim2006} while its study objective is to reduce transmission delay. Indeed, the bandwidth of ECG signal is much smaller than that of wireless communication signal, such as the cellular communication signal. Thus, within a wireless communication system, the transmission of ECG is not sensitive to the transmission delay. 

In the optimization of the bit rate, we calculate the minimum distortion between the ECG signal before and after the compression. The minimization is under the constraint of a given bit rate. The similar work is also performed in~\cite{Huang2017a}. However, the work in~\cite{Huang2017a} is limited in a theoretic view. The optimization in this paper is a practical solution. In the communication signal processing, the bit rate constraint optimization is also performed in~\cite{Huang2014a, Huang2014b,Huang2016,Huang2017}.


\section{ECG Compression Methods and Sensitivity to Beat Period Change}
\label{sec: seg_len}

We can easily observe that ECG samples are not independent, especially when they are close in time domain. Therefore, vector quantization is more efficient in compressing ECG signal than scalar quantization in usual cases. Among the diverse block signal transformations, Discrete cosine transform~(DCT) is considered in this paper due to its popularity and coefficients energy concentration. Afterwards, Huffman coding is applied in the DCT results to complete the vector quantization.



\subsection{Background of ECG Compression and Challenges}
\label{subsec: compression background}



Prior to DCT, ECG signal needs to be partitioned into segments. In this partition, there is a problem of the segment length selection. In equal partition, longer segments have smaller difference in the distribution of DCT results. The small difference means more conformance in codebook in the following Huffman coding which thus reduces the time in updating codebook. However, DCT on the longer segments costs more computation. In this subsection, we will show the tradeoff between the distribution stability and computation burden with respect to segment length. 


Let $x(t)$ denote ECG signal in continuous time domain, and $x[n]$ is its correspondent in discrete time domain. The ECG signal is divided into segments with the uniform length of $W$. The $m$-th ECG segment is denoted by $X^m$, that is,
\begin{equation}
X^m(w)=x[(m-1)W+w].
\label{eq: ecg seg def}
\end{equation}

Afterwards, the partitioned ECG segments are projected into an orthogonal space in the dimension of $W$. The projection is essentially a linear transform which is denoted by the multiplication of ECG vector $X^m$ with a transformation matrix $\Phi$ in the dimension of $m\times m$. The elements in the matrix $\Phi$ are the transformation coefficients. More specially, $\Phi$ is the coefficients in DCT. It is worth noting that $\Phi$ can also be the coefficients in other transforms, such as FFT, wavelet transformation. Transformation results are $C^m$ which are calculated by
\begin{equation}
C^m=X^m\Phi.
\label{eq: ecg seg def}
\end{equation}

The absolute value of the $m$-th element in $C^m(w)$ weights the projection of $X^m$ on the $w$-th row, $1\leq w\leq W$. The weights on all $W$ rows form a distribution. ECG signal is neither deterministic nor stationary. Thus, the elements $C^m$ for different $m$ are not constant, but have a distribution. 

Next, we quantitatively study the effect of segment length on the distributions. In the study, we compare the distribution via Kullback–Leibler divergence~(K-L) distance.

There are four steps towards calculating the K-L distance. First, we calculate the coefficients $C^m$ from the $m$-th ECG segment. Second, we calculate the distribution of the coefficients. Let $\mathcal{C}^m$ denote the set constituted by all distinguish elements of $C^m$, and $\nu_k$ denote the $k$-th element in $\mathcal{C}^m$, where $1\leq k\leq K$ and $K=|\mathcal{C}^m|$. The distribution is determined by the approximated PDF which is calculated by,
\begin{equation}
p^m(k)=\frac{|\{c|c=\nu_k, c\in C^m \}|}{|C^m|}.
\label{eq: PDF def}
\end{equation}


Third, we calculate K-L distance between two PDFs, say $p^{m_i}(w)$ and $p^{m_j}(w)$~\cite{Cover2006},
\begin{equation}
D_{KL}(i,j)=\sum_{k=1}^K p^{m_i}(k)\log\frac{p^{m_i}(k)}{p^{m_j}(k)},
\label{eq: KL dist}
\end{equation}
where $m_i, m_j\in\mathcal{M}$, $\mathcal{M}=\{1,2,\cdots,M\}$ and $m_i\neq m_j$. 

To comprehensively investigate the stability of the distributions, we build a set of $\mathcal{S}_W$ which elements are $D_{KL}(i,j)$, that is, $\mathcal{S}_W=\{D_{KL}(i,j)\}$ for $m_i,m_j\in\mathcal{M}$ and the subscript $W$ denotes the length of uniform window. Fourth, we calculate the variance of all elements in $\mathcal{S}_W$. The variance is denoted by $\sigma_W^2=var(\mathcal{S}_W)$. 

\begin{figure}[h]
	\vspace{0 mm}
	\centering
	\includegraphics[width=1\linewidth]{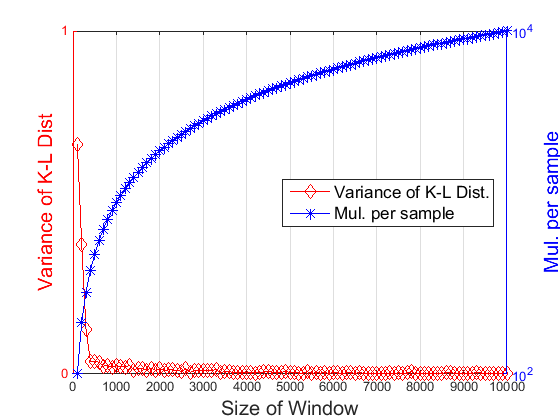}
	\caption{Variance of K-L distance versus window size and average number of multiplication per sample}
	\label{fig: KLD vs win}
	\vspace{0 mm}
\end{figure}

We select discretely valued $W$ and calculate the corresponding K-L distance variance $\sigma_W^2$. Fig.~\ref{fig: KLD vs win} plots $\sigma_W^2$ curve versus $W$ at the \textit{left} side \textit{y}-axis. In the plot, we use the ECG data from number 106 recording of MIT-BIH database.

The K-L distance variance curve indicates the variation of the weight distribution. For Huffman coding, the distribution determines the optimum codebook design. The smaller variance means less frequent codebook updating which thus reduce the computation burden. 

Besides the codebook design, DCT also a major computation expenditure. In DCT, multiplication consumes more hardware resources than addition operations. Thus, the average number of multiplications is taken as the metric for measuring the computation burden in DCT. At the length of $W$, we estimate the number of multiplications per ECG sample. The estimation results are plotted in the \textit{right} side \textit{y}-axis of Fig.~\ref{fig: KLD vs win}.


From Fig.~\ref{fig: KLD vs win}, we can observe that the variance of K-L distance decreases with window size increasing. From the observation, the distribution of DCT results changes significantly at small window size, while the variance of it approaches to zero at large window size. When the distribution have zero approaching variance, less computation is needed to update optimum compression codebook; otherwise, the compression scheme tends to recompute the optimum codebook to fit for the change of distribution. However, number of multiplications per ECG sample rises fast with increment of the window size. However, wearable devices for monitoring ECG signal are sensitive to the increment of computation burden. 

To break the tradeoff between the distribution stability and computation burden, we partition the ECG signal according to heart beat periods. After the beat period assisted partition, we compress ECG signal. The description of the ECG compression scheme is presented in the next subsection.

\subsection{Schematic Description on Proposed ECG Compression Scheme}
\label{subsec: system schematic}

Even though ECG signal is not strictly periodical or stationary, waveforms of ECG signal has similarities over any two heart beat periods. We call the heart beat period as {R-to-R period}. In the proposed compression scheme, we partition ECG signal based on R-to-R periods. After the R-to-R period assisted partition, the ECG segments have large similarity which means more stable distribution of DCT results $C^m$.

Thus, we need to detect R-waves first. In discrete time domain, let $n^m$ denote the time instance at which the $m$-th R-wave occurs and $n^{m+1}$ is for the $(m+1)$-th R wave. The ECG samples from $n^m$ to $(n^{m+1}-1)$-th time instance generate the $m$-th ECG segment, $X^m$, that is,
\begin{equation}
X^m=\{x[n^m],x[n^m+1],\cdots,x[n^{m+1}-1]\}^T. 
\label{eq: X m ecg}
\end{equation}

The distance between $m$- and $(m+1)$-th R wave is the $m$-th R-to-R period which is equal to the cardinality of $X^m$, $|X^m|$. This period is an important parameter for diagnosis and ECG signal reconstruction. Thus, besides the waveforms of ECG signal, the detected R-to-R period along with the locations of R-waves needs to be transmitted also. 

Next, we compress the partitioned ECG segments. In this paper, we consider two compression objectives, original ECG signal segment, and the differential ECG signal. In the compression, bit rate allocated for quantizing affects bandwidth consumption or size of occupied memory resource. In the mean time, the bit rate also affects the accuracy of reconstructed ECG signal from its quantization. Based on the relationship between the resource consumption and reconstruction accuracy, we optimize the compression scheme. In the optimization, we derive a practical bit rate allocation strategy by which reconstruction error is minimized at a given constraint of bit rate. 

\begin{figure}[!h]
\centering
\includegraphics[width=0.8\linewidth] {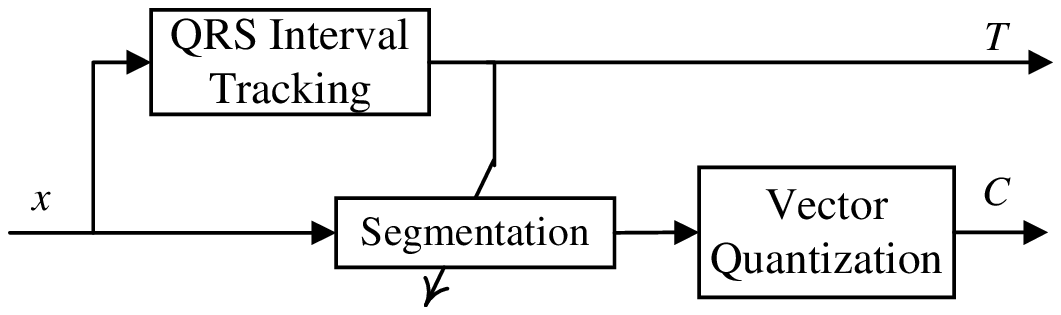}
\caption{General form of ECG compression scheme with R-R interval estimation assistance}
\label{fig: general form}
\end{figure}

To present an intuitive impression, the block diagram of our ECG compression scheme is plotted in Fig.~\ref{fig: general form} and the algorithm flow chart on high level is plotted in Algorithm~\ref{alg: algorithm compress}.

\begin{algorithm}
\caption{Data processing flow of ECG compression algorithm}
\label{alg: algorithm compress}
\begin{algorithmic}[1]
\State Detect first two R-waves;
\For{full ECG signal length}
\State Track R-waves;
\State Partition ECG signal into the segments at R-waves;
\State Normalize the length of ECG segments;
\State Optimize bit rate and quantize ECG segments;
\State Send quantization results to the receiver.
  \EndFor
\end{algorithmic}
\end{algorithm}

From Fig.~\ref{fig: general form} and Algorithm~\ref{alg: algorithm compress}, there are three modules in the optimum ECG compression scheme. The R-R interval tracking component is responsible for continuously detecting R-waves. With the detected R-waves, segmentation module divides ECG signal into segments at the detected R-waves. For uniforming segment length, the interpolation and decimation are performed in the segmentation module. Optimization of bit rate and codebook design are performed in Vector Quantization module.

\section{R-waves Detection and Tracking}
\label{sec: QRS detect}

From the introduction in the previous section, R-waves detection is needed before the R-to-R ECG data partition. In this section, we present a new method of detecting R-waves.

The ECG waveforms are usually different from person to person. Even for the same person, ECG signals acquired from different electrodes are also different. Take a standard ECG system for example. Within the lead \textrm{I} ECG signal, R-wave is usually higher than T-wave in amplitude. For the signal from lead V4 or V5, T-wave might be higher than R-wave in a large probability. Due to the variety of ECG waveforms, it is difficult to build a R-wave detector in a general form. 

Even though waveforms of ECG signals are significantly different, there is a common feature in ECG signals, that is, ECG signal changes faster within QRS duration than that in other part of a R-to-R period. This feature can be utilized to construct a R-wave detector in a general form. 

The faster changing feature in QRS duration can be observed from order one differential ECG signal. However, the change rate are not the same at different ECG recordings. For example, differential QRS segment has higher amplitude in the positive part for some ECG recordings, while the negative part might be stronger for other recordings. To comprise the diverse ECG recordings, we need a further processing on the differential ECG signal such that R-wave can be detected in a general form. No matter in what shape a differential ECG waveform is, a QRS segment possesses the strongest energy after the differentiation. 

We employ TK operator to detect the energy in the differential ECG waveform since a waveform in a uniform shape can be obtained after the processing by the TK operator. The uniform shape is beneficial for implementing R-wave detection in a general form. Along with the path of 'differentiation-energy detection by TK operator', we propose a novel R-wave detector.

\subsection{TK Operator Based Preprocessing}
\label{subsec: preprocess}

In the subsection, we introduce the details of how to convert diverse ECG signals into waveforms in a uniform shape. The conversion is called as the preprocessing before R-to-R period estimation. 

There are three steps to complete the preprocessing. First, we remove baselines and noise from raw ECG signals. Second, order one differential ECG signal is calculated. Third, a Teager-Kaiser operator is taken to calculate the energy of the differential ECG signal. 

To avoid trivial description, the details of baseline removal will not be presented. Let $x[n]$, $n\in\mathbb{Z}$, denote the ECG signal after the baseline removal. $\Delta x$ denotes the differential ECG signal which is calculated by subtracting ECG signal with its replica with one sample delay, 
\begin{equation}
\Delta x[n]=x[n+1]-x[n].
\label{eq: diff ecg cal}
\end{equation}

 The order one differential ECG is used due to the following three reasons. 

\begin{enumerate}
\item QRS segments possess stronger energy at differential ECG signals which is caused by the fast change of ECG signal within the QRS segments. The other waves, such as T-, P-waves, may have the higher amplitude while their changing rate is usually much slower.

\item The feature of possessing stronger power by QRS segments widely exists. For the same person, QRS segments has stronger power for all leads in a standard ECG system. For different persons, QRS segments still have the stronger power in a differential ECG form.

\item Differential ECG signal in higher order ($\geq 2$) does not have the common features within QRS segments.

\end{enumerate}

Even though differential ECG signals fluctuates at a stronger amplitude within QRS segments, the waveforms of the differential ECG signals are not in a common shape. We employ Teager-Kaiser~(TK) operator to convert the differential ECG signal into ones in a similar shape. The application of the TK operator on $\Delta x$ is presented as follows,
\begin{equation}
y[n]=(\Delta x[n])^2-\Delta x[n-1]\Delta x[n+1].
\label{eq: TK diff ecg cal}
\end{equation}

\begin{figure}[h!]
\centering
\includegraphics[width=1\linewidth] {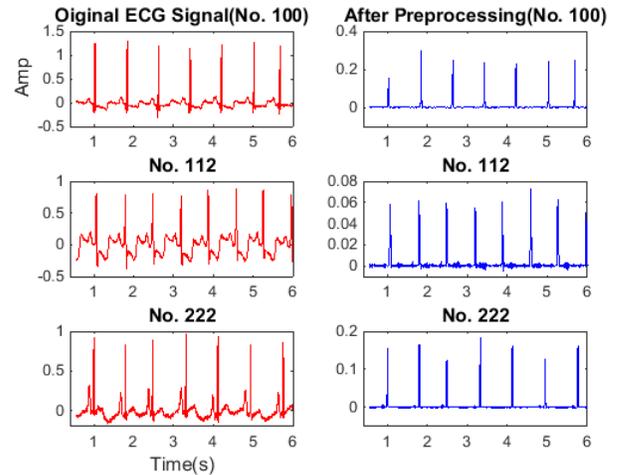}
\setlength{\abovecaptionskip}{3pt plus 3pt minus 2pt}
\caption{Demo of preprocessing ECG signal}
\label{fig: preprocessing effect}
\end{figure}

We first show the effectiveness of proposed method in uniforming the ECG waveforms. In the left column of Fig.~\ref{fig: preprocessing effect}, the number 100, 112, and 222 ECG recordings in MIT-BIH database are plotted. The preprocessing results are plotted at the right column of Fig.~\ref{fig: preprocessing effect}.

 From Fig.~\ref{fig: preprocessing effect}, the proposed preprocessing method enables us obtain signal in a regular and common form regardless of the variety of original ECG waveform. We need also to notice that the preprocessing results retains the information of R-to-R periods. Next, we will introduce a method of R-to-R period estimation based on the preprocessing results.

\subsection{A Novel R-wave Detection}
\label{subsec: QRS detection}

From Section~\ref{subsec: preprocess}, the preprocessing results are in spike-like shape where the spikes occur within QRS segments. Once we detect the spikes, the locations of R-waves are correspondingly obtained.

\begin{figure}[!h]
\centering
\includegraphics[width=0.6\linewidth] {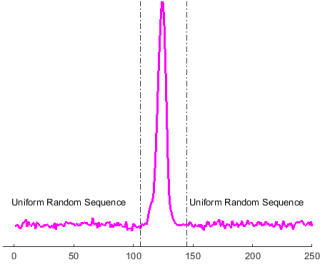}
\caption{First order differential ECG template for QRS complex detection}
\label{fig: QRS mdoel}
\end{figure}

To detect the spikes, we select a template describing the general shape of the spikes. The template is denoted by $\bold{s}$ as shown in Fig.~\ref{fig: QRS mdoel}. The template is obtained via averaging TK operator output from ten ECG recordings in MIT-BIH database. In Fig.~\ref{fig: QRS mdoel}, the central spike is the result of strong energy concentrating in QRS segments. The two sides of the template are uniformly distributed random sequences with zero mean. 

We calculate the similarity of $\bold{s}$ and $y$ to detect the spikes. The similarity is measured by the correlation between $\bold{s}$ and $y$. A spike is regarded to be existing when the correlation is larger than a threshold. The correlation is calculated by 
\begin{equation}
\kappa[n_{\tau}]=\frac{1}{N}\sum_{n=1}^N s[n]y[n-n_{\tau}],
\label{eq: correlation def}
\end{equation}
where $N$ is the number of ECG samples which is equal to the length of the template $\bold{s}$; $\kappa$ is the correlation function.

Without loss of generality, we consider a sampling frequency of 365 Hz which is adopted in MIT-BIH ECG database. To detect the first R-wave, we take two consecutive segments of ECG data, $\bold{y}_1$ and $\bold{y}_2$, each with the length of 256, $\bold{y}_1=\{y[n+1], y[n+2],\cdots, y[n+N]\}$ and $\bold{y}_2=\{y[n+N+1], y[n+N+2],\cdots, y[n+2N]\}$, $N=256$. At the sampling frequency of 365 Hz, there are usually no two R-waves within the duration of 256 ECG samples and there will be one R-wave within the duration of 512 samples. 

At the two pairs of data sets $(s, \bold{y}_1)$ and $(s, \bold{y}_2)$, we calculate the correlation respectively which results are denoted by $\bold{\kappa}_1$ and $\bold{\kappa}_2$, $\bold{\kappa}_1=s\otimes\bold{y}_1$, $\bold{\kappa}_2=s\otimes\bold{y}_2$ where {$\otimes$} denotes the operation of correlation. Afterwards, we select the maximum from the set $\bold{\kappa}_1 \cup \bold{\kappa}_2$. The selected maximum value informs us the location of the differential ECG spike. Until now, the principle of how to detect the first spike in the differential ECG is introduced. Next, the details of implementing the detection will be introduced.

 We select FFT as the tool to implement the detection of the first spike since FFT is widely used and there are plenty of off-the-shelf IC chips. Furthermore, FFT is able to speed up the computation in the detection. Algorithm~\ref{alg: FFT QRS detect} presents the novel R-wave detection. 

\begin{algorithm}
\caption{Proposed R-wave detection}
\label{alg: FFT QRS detect}
\begin{algorithmic}[1]
\State As shown in (\ref{eq: TK diff ecg cal}), apply the TK operator on the ECG signal $x$ to obtain $y$;
\State From arbitrarily selected a sample in $y$, abstract two adjacent data blocks, $\bold{y}_1$ and $\bold{y}_2$, each with 256 samples;
\State Calculate Fourier transform of $\bold{s}$, $\bold{y}_1$ and $\bold{y}_2$,
\begin{equation}
\begin{aligned}
S_s&=\mathcal{FFT}\{\bold{s}\}\\
S_{y_1}&=\mathcal{FFT}\{\bold{y}_1\},
S_{y_2}=\mathcal{FFT}\{\bold{y}_2\}.
\end{aligned}
\label{eq: FFT corr}
\end{equation}
\State Calculate the dot product between $S_s$ and $S_{y_1}^*$, and the same for $S_s$ and $S_{y_2}^*$,
\begin{equation}
\begin{aligned}
Z_1=S_s\cdot S_{y_1}^*,  Z_2=S_s\cdot S_{y_2}^*,
\end{aligned}
\label{eq: FFT dot prod}
\end{equation}
where $(\cdot)^*$ denotes the conjugation operation. 
\State Calculate inverse Fourier transform of $Z_1$ and $Z_2$, 
\begin{equation}
\bold{\kappa}_1=\mathcal{IFFT}\{Z_1\}, \bold{\kappa}_2=\mathcal{IFFT}\{Z_2\}.
\label{eq: FFT dot prod}
\end{equation}
\State Select the maximum point from $\bold{\kappa}_1\cup\bold{\kappa}_2$ which is denoted by $\kappa_M$. The position index of $\kappa_M$ in $\bold{\kappa}_1\cup\bold{\kappa}_2$ informs us the location of the QRS wave. 
\end{algorithmic}
\end{algorithm}

As for Algorithm~\ref{alg: FFT QRS detect}, there is a notation worth mention. Since the template $\bold{s}$ is priorly known, Fourier transform of $\bold{s}$ can be stored in memory ahead of the computation. The pre-storage of Fourier transform of $\bold{s}$ is able to alleviate the burden in each round of computation. Furthermore, the stored values can be taken as constants. Thus, the dot multiplication in (\ref{eq: FFT dot prod}) can be implemented via bit shifting which is in low hardware cost, not by multiplier, such as DSP48s, which are usually precious in chips. 

\textbf{Computation complexity:} In R-wave detection, FFT is used due to its advantage in computation saving. As a well known result, numbers of both multiplications and additions in $N$-scale FFT are on the scale of $\mathcal{O}(N\log N)$. Thus, we need $(4\times2048+2\times256)$ multiplications in Algorithm~\ref{alg: FFT QRS detect}, and the same number with the addition. If FFT is not utilized in the computation, $2\times65536$ multiplications are needed.

\section{Frame Level Bit Rate Allocation}
\label{sec: bit rate allocation}

According to the description of the proposed ECG compression scheme in Section~\ref{subsec: system schematic}, the detection of R-waves are beneficial for achieving more stable statistical feature between ECG segments. The stability reduces the burden for calculating an optimum compression scheme in the ECG signal transmission. In this section, we will show how to calculate the optimum scheme.

\subsection{Principle of rate-distortion function guided optimum bit rate allocation}
\label{subsec: opt bit allocation}

Rate distortion theory is the fundamental in the optimization of lossy quantization. We borrow the idea in rate distortion theory to design a practical optimum compression scheme. The compression scheme is built in the framework of combining DCT and Huffman coding. Within the compression scheme, we calculate the minimum compression loss with a bit rate constraint. 



In the compression scheme, DCT is first applied on ECG segments. Afterwards, DCT results are quantized, and the quantization results are then encoded using Huffman coding method. In this section, we optimize the quantization such that the compression loss is minimized under a constraint of bit rate. 



\subsubsection{Calculation of compression loss}

$X^m$ is the $m$-th ECG segment after R-to-R period based partition. Let $C^m$ denote DCT of $X^m$. Afterwards, Huffman coding is applied on $C^m$. After the encoding, the codeword is denoted by $\psi^m$. The codebook is $\Psi^m$.  Let $\iota^m$ denote the length of the codewords $\psi^m$, $\iota^m=\sum |\psi^m|$, where $\psi^m\in\Psi^m$.

At the receiver side, $\psi^m$ is decoded first and the decoding results are the DCT coefficients $\hat{C}^m$. From $\hat{C}^m$, we can further reconstruct the $m$-th ECG segment $\hat{X}^m$ via inverse DCT.

Let $d^m$ denote the mean square error~(MSE) between $X^m$ and $\hat{X}^m$. The MSE is taken to measure the reconstruction accuracy. $d^m$ is calculated by
\begin{equation}
\begin{aligned}
d^m=&E[\sum_{w=1}^W(X^m(w)-\hat{X}^m(w))^2]\\
=&E[\parallel X^m- \hat{X}^m\parallel^2]\\
=&E[\parallel\Phi^{-1}\parallel\cdot\parallel \Phi(X^m)- \Phi(\hat{X}^m)\parallel^2]\\
=&E[\parallel\Phi^{-1}\parallel\cdot\parallel C^m- \hat{C}^m\parallel^2].\\
\end{aligned}
\label{eq: distortion def}
\end{equation}

For DCT, $\parallel\Phi^{-1}\parallel$ in (\ref{eq: distortion def}) is a constant. Therefore, the reconstruction error is determined by the mean square error between DCT coefficients $C^m$ and the its reconstruction $\hat{C}^m$. 

The DCT result $C^m$ is a vector with elements in real numbers. In a practical system, we need to quantize $C^m$ before the Huffman coding since discrete elements with probability support are needed by the Huffman encoder. Huffman coding is an entropy coding which does not induce compression loss. Therefore, the MSE between $C^m$ and $\hat{C}^m$ is fully determined by the loss in quantizing $C^m$. 


Let $p_c(c)$ denote the distribution of $C^m$, where $c\in C^m$. In the quantization of $C^m$, $i$-th partition zone in the value range of $c$ is denoted by $[c_l, c_{l+1})$ where $i\in\{0,1,\cdots,L\}$ and $L$ is equal to the number of quantization levels. The DCT results within the zone $[c_l, c_{l+1})$ are mapped to $\hat{c}_l$. Since the differences between $C^m$ and $\hat{C}^m$ are caused by the quantization, we have 
\begin{equation}
\parallel C^m- \hat{C}^m\parallel^2= \sum_{l=1}^L\int_{c_l}^{c_{l+1}}(c_i-\hat{c}_i)^2p_c(c)dc.
\label{eq: diff between Cand C hat}
\end{equation}

Substituting (\ref{eq: diff between Cand C hat}) into (\ref{eq: distortion def}), we have 
\begin{equation}
d^m=\parallel\Phi^{-1}\parallel \sum_{l=1}^L\int_{c_l}^{c_{l+1}}(c_i-\hat{c}_i)^2p_c(c)dc.
\label{eq: distortion def 2}
\end{equation}

\subsubsection{Optimum quantization under a bit rate constraint}
In the optimization of ECG compression, we minimize the MSE $d^m$ under a constraint of bit rate. Bit rate is determined by the number of bits used for quantizing ECG segments. 

For Huffman coding, the bit rate is fully determined by the distribution of the quantized results $\{\hat{c}_l\}$, $l\in\mathcal{L}$. The distribution on  $\{\hat{c}_l\}$ is denoted by $p_{\hat{c}}(\hat{c}_l)$. For the known Borel set $\{c\}$ and the distribution $p_c(c)$ defined on $\{c\}$, the quantization on $\{c\}$ determines the distribution $p_{\hat{c}}(\hat{c}_l)$.



 Let $\zeta$ denote the codewords from the Huffman encoder and $\nu$ denote the average length of the codewords, $\nu=|\zeta|$. As a well known result, the average length of codewords in a codebook is no smaller than entropy of the source. Thus, we have
\begin{equation}
\nu\geq\sum_{l=0}^L p_{\hat{c}}(\hat{c}_l)\log\frac{1}{p_{\hat{c}}(\hat{c}_l)},
\label{eq: ave len Huffman}
\end{equation}
where the right side of (\ref{eq: ave len Huffman}) follows the definition of entropy for a discrete source, and
\begin{equation}
p_{\hat{c}}(\hat{c}_l)=\int_{c_l}^{c_{l+1}}p_c(c)dc.
\label{eq: p hat c def}
\end{equation}

In the transmit of the $m$-th ECG segment, we set an upper bound on the bit rate which is the constraint in the optimization. The upper bound is denoted by $R$. The average length of codewords is not larger than the upper bound, that is, 
\begin{equation}
\nu\leq R.
\label{eq: bit constraint}
\end{equation}

From (\ref{eq: ave len Huffman}) and (\ref{eq: bit constraint}), we have 
\begin{equation}
\sum_{l=0}^L p_{\hat{c}}(\hat{c}_l)\log\frac{1}{p_{\hat{c}}(\hat{c}_l)}\leq R.
\label{eq: bit constraint 2}
\end{equation}

Until now, we already define the objective function and the constraint in the optimization problem. Next, we introduce the solution of the problem.

\subsubsection{Calculation of optimum compression scheme}

Remember that our goal is to minimize the distortion $d_m$ under the bit rate constraint of (\ref{eq: bit constraint 2}). The constrained minimization is explicitly written as follows,
\begin{equation}
\left\{
\begin{aligned}
&\min_{c_l:l\in\mathcal{L}\cup\{0\},\hat{c}_l:l\in\mathcal{L}}  d_m\\
&\sum_{l=0}^L p_{\hat{c}}(\hat{c}_l)\log\frac{1}{p_{\hat{c}}(\hat{c}_l)}\leq R
\end{aligned}
\right..
\label{eq: constrained opt}
\end{equation}

The constrained optimization in (\ref{eq: constrained opt}) is converted into an unconstrained one. Thus, we have
\begin{equation}
\begin{aligned}
J=&d_m+\lambda\cdot\nu\\
=&\parallel\Phi^{-1}\parallel \sum_{l=1}^L\int_{c_l}^{c_{l+1}}(c_l-\hat{c}_l)^2p_c(c)dc\\
&+\lambda\cdot\sum_{l=0}^L p_{\hat{c}}(\hat{c}_l)\log\frac{1}{p_{\hat{c}}(\hat{c}_l)},
\label{eq: unconstrained opt}
\end{aligned}
\end{equation}
where $\lambda$ is a Lagrange constant.

We borrow the idea of iterative descent algorithm to solve the optimization problem (\ref{eq: unconstrained opt}). In the optimization, there are two types variables needs to be determined, the segmentation of quantization $c_l$ and the quantized output $\hat{c}_l$ . The former determines the average bit rate and the latter affects the reconstruction accuracy. 

To optimize the segmentation in the quantization, we calculate the zero valued first order derivative of $J$ with respect to $c_l$. The derivations are presented as follow,
\begin{equation}
\begin{aligned}
\frac{\partial J}{\partial c_l}=&2\parallel\Phi^{-1}\parallel (c_l-\hat{c}_l)p_c(c_l)+2\parallel\Phi^{-1}\parallel (c_l-\hat{c}_{l-1})p_c(c_l)\\
+&p_c(c_l)\left(2+\log\int_{c_l}^{c_{l+1}}p_c(c)dc+\log\int_{c_{l-1}}^{c_{l}}p_c(c)dc\right)\\
\cong&2\parallel\Phi^{-1}\parallel (c_l-\hat{c}_l)p_c(c_l)+2\parallel\Phi^{-1}\parallel (c_l-\hat{c}_{l-1})p_c(c_l)\\
+&p_c(c_l)\left(2+\log p_{\hat c}(\hat{c}_l)+\log p_{\hat c}(\hat{c}_{l-1})\right)\\
=&0.
\label{eq: derivative over h l}
\end{aligned}
\end{equation}

Solving (\ref{eq: unconstrained opt}), we have
\begin{equation}
c_l=\frac{1}{2}\left(\frac{2+\log p_{\hat c}(\hat{c}_l)+\log p_{\hat c}(\hat{c}_{l-1})}{2\parallel\Phi^{-1}\parallel}+\hat{c}_l+\hat{c}_{l+1}\right)\\
\label{eq: cal of x l}
\end{equation}

To determine quantization outputs, we calculate the derivative of $J$ with respect to $\hat{c}_l$, that is,
\begin{equation}
\begin{aligned}
\frac{\partial J}{\partial \hat{c}_l}=&\parallel\Phi^{-1}\parallel\int_{c_l}^{c_{l+1}}2(c-\hat{c}_l)p_c(c)dc\\
=&0.
\label{eq: derivative over hat h l}
\end{aligned}
\end{equation}

From (\ref{eq: derivative over hat h l}), $\hat{c}_l$ is calculated by
\begin{equation}
\begin{aligned}
\hat{c}_l=&\frac{\int_{c_l}^{c_{l+1}}cp_c(c)dc}{\int_{c_l}^{c_{l+1}}p_c(c)dc}\\
\label{eq: solu derivative over hat h l}
\end{aligned}
\end{equation}

The two equations (\ref{eq: cal of x l}) and (\ref{eq: derivative over hat h l}) provide us the condition at which we are able to achieve the minimum $d_m$. Along with the idea of iterative descent algorithm, we present a practical method of approaching the optimization.


Initially, we divide the domain of DCT result into $L+1$ zones with uniform duration. In each zone, we select one point as the initial quantization output. Next, we iteratively calculate the following steps to approach the optimum condition, 
\begin{enumerate}
\item Calculate the quantization codeword $\hat{c}_l$ via (\ref{eq: derivative over hat h l}).
\item Calculate the probability of the corresponding segment $[c_l, c_{l+1})$ which contains $\hat{c}_l$.
\item Calculate the ending points of each segment via (\ref{eq: cal of x l}).
\item $J$ is calculated via (\ref{eq: unconstrained opt}).
\item Judge whether the iterative computation has already met the stopping condition, $\frac{|J^{(i)}-J^{(i-1)}|}{J^{(i)}}<\xi$. If the condition is not satisfied, the iterative computation will continue from step 1.
\end{enumerate}

From the iterative optimization, we obtain an optimum bit rate and distortion (MSE)  pair $D_L-R_L$. Remember that the subscript $L$ denotes the number of discrete levels used for quantizing $\{c\}$. We change the value of $L$ and repeat the calculation of $D_L-R_L$ curves. Thus, we obtain a group of $D_L-R_L$ curves at different values of $L$. Next, we calculate the convex of the group of $D_L-R_L$ curves for all $L$. The convex is denoted by a $D-R$ curve. 

With the $D-R$ curve, we know the minimum distortion in the condition of $L$-level quantization of DCT coefficients and its achieving conditions. The conditions include the boundaries of quantization zones and the quantization output in each zone.

 For example, we first set a budget of bit rate $R$. With the budget, we know the minimum distortion $\min{d_m}=D$ from the calculated $D-R$ curve. Besides that, we also know the value of $L$ at which the minimum distortion is achieved. Furthermore, we derive the condition of how to quantize the DCT result for achieving the minimum distortion with the algorithm above.

\section{Frame Level ECG Compression}
\label{sec: ecg compression}

Until now, the main steps in the algorithm of compressing ECG signals have been introduced. In this section, we discuss the concrete structures based on which we can implement the algorithm.

\begin{figure*}[h]
\centering
\includegraphics[width=0.8\linewidth] {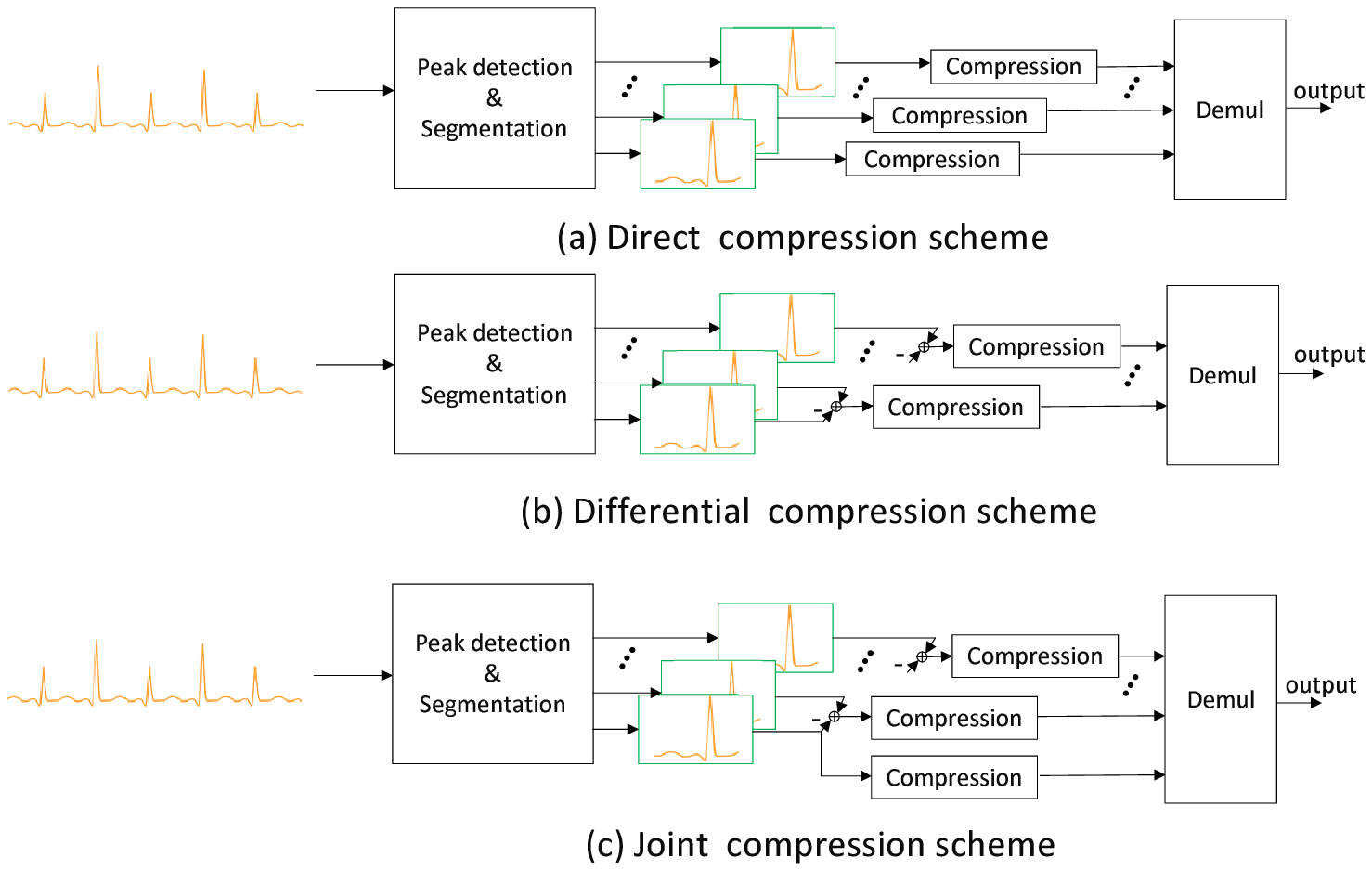}
\caption{High level description of the three compression structures}
\label{fig: high level description}
\end{figure*}

We consider three structures. All the three compression schemes use the common R-peak detection method, which is proposed in this paper, to segment the ECG signal. After the segmentation, different compression schemes are adopted in the three structures. For assist our explanation, the high-level descriptions of the three compression structures are plotted in Fig.~\ref{fig: high level description}. In the first scheme, original ECG segments are directly compressed and transmitted, as shown in Fig.~\ref{fig: high level description}(a). Fig.~\ref{fig: high level description}(b) shows the second structure in which we calculate the difference between the two adjacent ECG segments. Then, the difference is compressed for transmission. The combination of previous two is the third structure which is called as the joint compression, as shown Fig.~\ref{fig: high level description}(c). In third structure, the ECG segments are grouped first. Then, the first segment among each group is directly compressed, and the other segments within the group are differentially quantized for transmission. 

\subsection{Compression of original ECG signal}
\label{subsec: original ECG}

Using the method introduced in Section~\ref{sec: QRS detect}, R-waves are first detected. Let $\Omega$ denote the set of R-wave locations, that is, $\Omega=\{n^R_i\}$, $n\in\mathbb{Z}$, where $n^R_i$ denotes the location of $i$-th R-wave. Furthermore, let $X^i$ denote the set of all ordered ECG samples between $(n^R_i+1)$ and $(n^R_{i+1})$, $X^i=\{x[n^R_i+1], x[n^R_i+2],\cdots, x[n^R_{i+1}]\}$. Generally, the cardinality of $X^i$ at different $i$ changes. 

\begin{figure*}[h]
\centering
\includegraphics[width=1\linewidth] {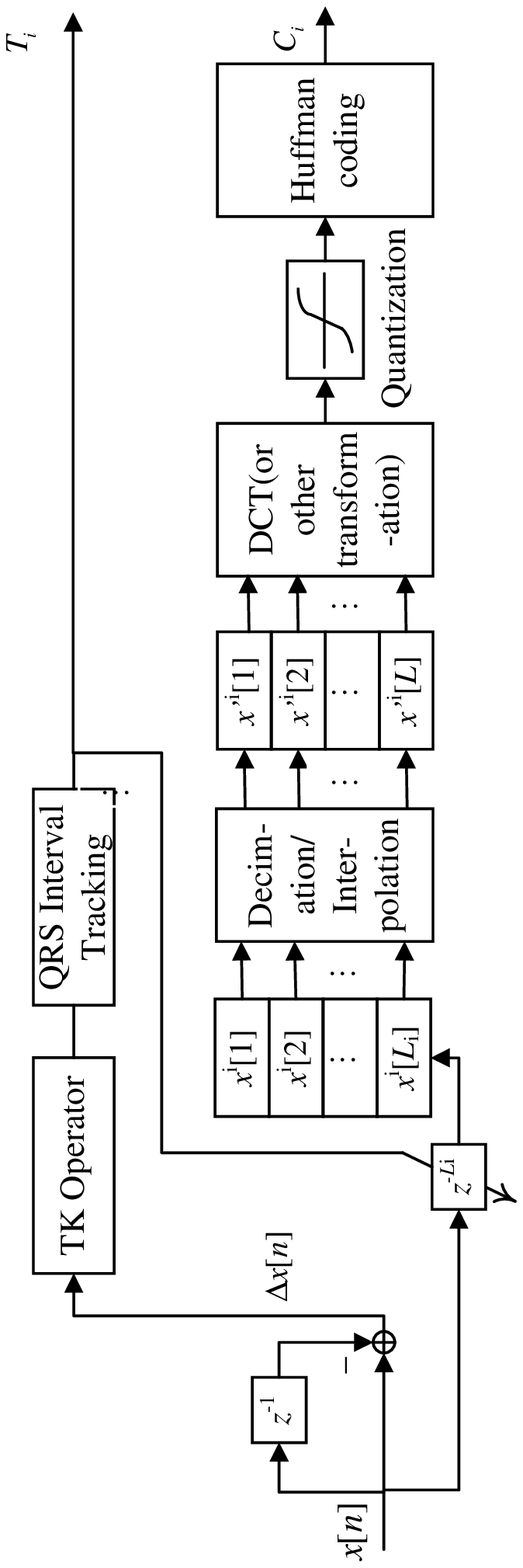}
\caption{Direct ECG compression scheme for transmission}
\label{fig: original ecg scheme}
\end{figure*}

For processing convenience, we convert all segments $X^i$ into the ones with the uniform length of $L$. Let $X^i$ denote the conversion result and $X^{\prime i}=\{x^{\prime}_i[1], x^{\prime}_i[2],\cdots, x^{\prime}_i[L]\}$. Afterwards, DCT is performed on $X^{\prime i}$ and we obtain the DCT coefficients $C^i$. Then, $C^i$ is quantized and Huffman encoded. The bit rate used in the quantization and encoding is determined by the method in Section~\ref{sec: bit rate allocation}. After the encoding, codewords and the $i$-th R-to-R period $T_i$ are transmitted. To present an intuitive impression, the block diagram of the direct compression scheme is shown in Fig.~\ref{fig: original ecg scheme}. 

At the receiver side, the codewords are decoded by Huffman decoder following which is inverse DCT. Right now, the output of the inverse DCT are ECG segments in a uniform length. Via decimation or interpolation, the ECG segments in a uniform length are then converted into the ones with original length of $T_i$. The conversion result is denoted by $\hat{X}^i$.

The direct compression scheme is able to be implemented using a simple structure. However, higher bit rate is needed since original ECG signal usually has larger dynamic range than the differential ECG signal does. The high demand on bits will be shown in experiment section.
 
 
\subsection{Compression of differential ECG signal}
\label{subsec: diff ECG}

Differential ECG signal has a smaller dynamical range which allows the similarly accurate compression at less quantization levels. Besides that, Differential ECG signal has more stable statistic than original one.

\begin{figure*}[h]
\centering
\includegraphics[width=1\linewidth] {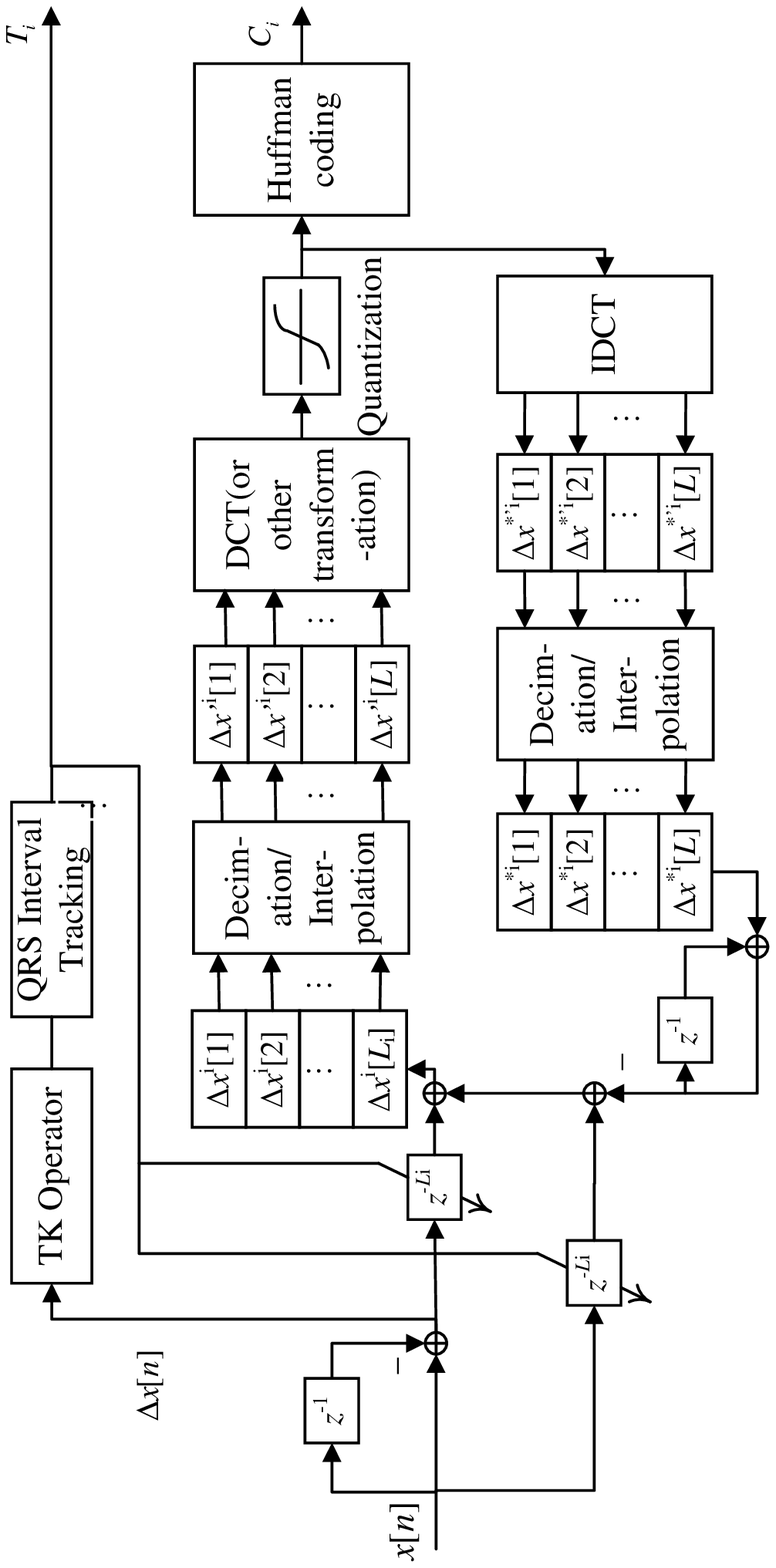}
\caption{Differential ECG compression scheme}
\label{fig: diff ecg scheme}
\end{figure*}

Fig.~\ref{fig: diff ecg scheme} presents the block diagram of the differential ECG compression scheme. After the R-wave detection, we partition ECG signal into segments based on R-to-R periods. Same with the notations in the previous subsection, set of R-wave locations is denoted by $\Omega$ and $\Omega=\{n^R_i\}$. 

Next, we calculate the differential ECG signal. Let $\Delta X^i$ denote the $i$-th differential ECG segment where $\Delta X^i=\{\Delta x[n^R_i], \Delta x[n^R_i+1],\cdots, \Delta x[n^R_{i+1}-1]\}$ and $\Delta x[n^R_i+j]=x[n^R_i+j+1]-x[n^R_i+j]$, $0\leq j\leq n^R_{i+1}-n^R_i-1$. Via interpolation/decimation, we obtain the differential ECG signal in uniform length which is denoted by $\Delta X^{\prime i}$, where  $\Delta X^{\prime i}=\{\Delta x^{\prime}_i[1], \Delta x^{\prime}_i[2],\cdots, \Delta x^{\prime}_i[L]\}$. The following operations in compressing the differential ECG segments are the same with the ones in the direct compression structure.  

Comparing Fig.~\ref{fig: original ecg scheme} with Fig.~\ref{fig: diff ecg scheme}, the compression of differential ECG signal needs more computations. The increased computations are mainly induced by calculating differential ECG signal. On the other side, the compression of differential ECG signal is able to more effectively save bit rate due to its smaller dynamic range and stabler statistical feature.

\subsection{Compression of joint ECG signal}
\label{subsec: joint ECG}

\begin{figure*}[!h]
\centering
\includegraphics[width=1\linewidth] {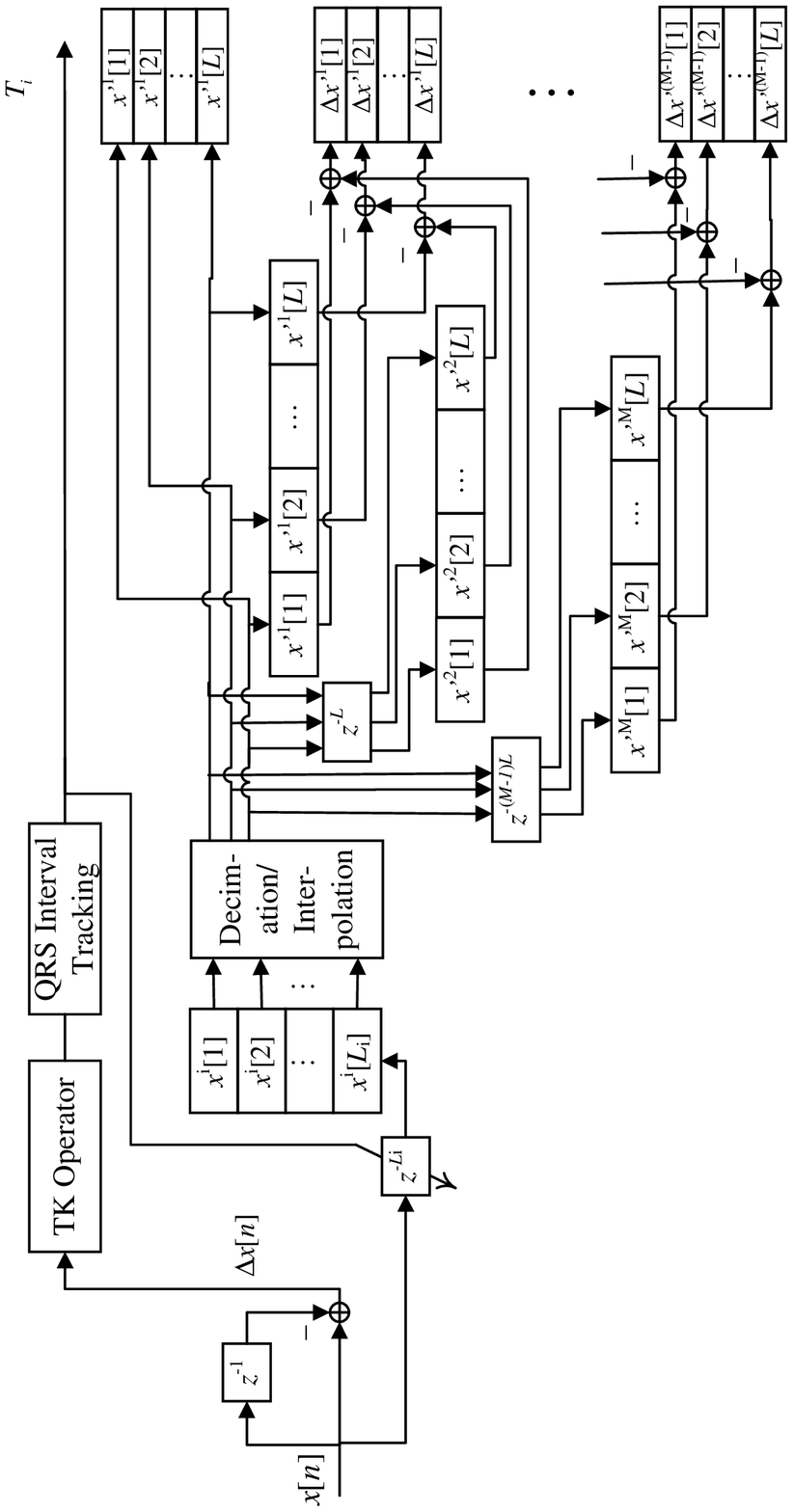}
\caption{Joint ECG transmission scheme}
\label{fig: joint ecg scheme}
\end{figure*}

In this subsection, we consider a joint ECG compression structure as shown in Fig.~\ref{fig: joint ecg scheme}. In the joint ECG compression structure, both the original and differential ECG signals are compressed and transmitted. In the joint scheme, $M$ ECG segments constitute a group. The $j$-th group, $G_j$, is taken as an example in our analysis where $G_j$ is written as follows, 
\begin{equation}
G_j=X^{(j-1)\times M+1}\cup X^{(j-1)\times M+2}\cup\cdots\cup X^{(j-1)\times M}. 
\label{eq: def group}
\end{equation}

In the joint compression scheme, we first normalize the length of all elements in the $G_j$ via interpolation or decimation. After the length normalization, we obtain a new group $G_j^{\prime}$, 
\begin{equation}
G_j^{\prime}=X^{\prime (j-1)\times M+1}\cup X^{\prime (j-1)\times M+2}\cup\cdots\cup X^{\prime j\times M},
\label{eq: def group same len}
\end{equation}
where the sizes of $X^{\prime j\times M+m}$'s are all equal to $L$ where $1\leq m\leq M$.

Within the $j$-th group $G_j^{\prime}$, we calculate the difference between adjacent segments, 
\begin{equation}
\Delta X^{\prime m}_j=X^{\prime(j-1)\times M+m+1}-X^{\prime (j-1)\times M+m},
\label{eq: diff set}
\end{equation}
where $1\leq m\leq(M-1)$.

Afterwards, within the $j$-th group, the first segment $X^{\prime (j-1)\times M+1}$ and all $(m-1)$ differential ECG segments ${\Delta X^{\prime m}_j}$, $1\leq m\leq(M-1)$, are compressed for transmission. Compared with the differential compression structure, the joint structure does not require the feedback loop for correcting the error since the first segment in each group is transmitted for periodically correcting the error. The correction, served by the first ECG segment, is able to avoid the accumulation of differentiation errors. 



\section{Experimental Verifications}
\label{sec: experiment}

To validate the proposed ECG compression structure, experiments are performed in this section. In the experiment, the data are from MIT-BIH Arrhythmia database~\cite{Goldberger2000}. The ECG data are sampled at the frequency of 360 Hz and quantized by 11 bits. 

We perform the tests in four aspects. First, we illustrate that R-to-R period based partition is beneficial for reducing computation burden in codebook design. Second, we compare compression methods with and without bit rate optimization. Next, we test the three compression structures in Section~\ref{sec: ecg compression}. Finally, we compare the performance of our ECG compression scheme with the ones in literature.



\begin{figure}[h]
\centering
\includegraphics[width=1\linewidth] {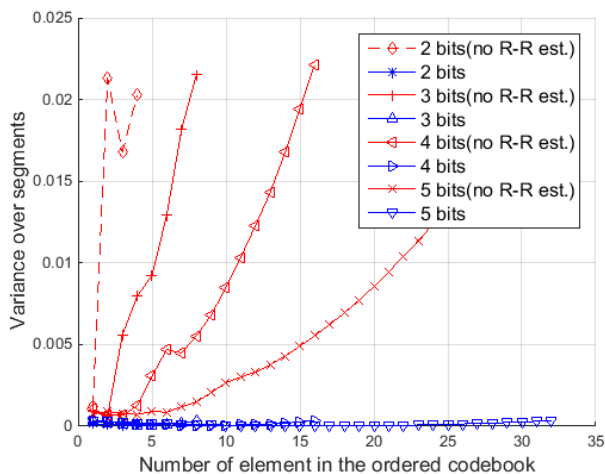}
\setlength{\abovecaptionskip}{3pt plus 3pt minus 2pt}
\caption{Variance comparison of codeword in optimum codebook}
\label{fig: codebook comparison}
\end{figure}

As analyzed before, ECG segments obtained from R-to-R period based partition have large similarity between each other. Since an optimum codebook is related to the distribution of ECG samples in each segment, the larger inter-similarity between segments means more stable optimum codebook. The stability of the codebook means less computation in the optimization. We calculate the optimum codebooks on all ECG segments. Among the codebooks, we calculate the variance of each codewords within the codebooks. The smaller variance means the smaller change of the codeword. As comparison, even-length partition is also performed in the ECG signal. The calculation of the optimum codebook and codeword variance is also performed on the even-length partitioned segments. Fig.~\ref{fig: codebook comparison} plots the variance curves at each codeword.






From Fig.~\ref{fig: codebook comparison}, the advantage of R-R period estimation assisted codebook optimization can be easily observed. First, with the assistance of R-wave detection, we obtain the codebooks in a smaller diversity than those without R-wave detection. 

We can also observe that the variance of codewords in even-length partition becomes larger with the increase of quantization levels. In the meantime, the variance derived from our new method is stable. The stability enables us to save the time and computation in designing codebooks since there is small change of codebook between ECG segments. For the even-length partition, the calculation of codebook optimization needs to be performed more frequently; otherwise, the compression loss becomes large.  

\begin{figure}[h]
\centering
\includegraphics[width=1\linewidth] {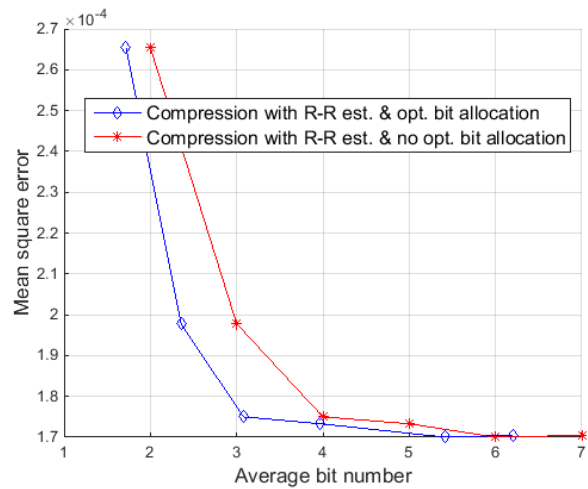}
\setlength{\abovecaptionskip}{3pt plus 3pt minus 2pt}
\caption{Bit rate allocation optimization in ECG compression with R-R interval estimation}
\label{fig: bit rate opt effect}
\end{figure}

We also investigate the effectiveness of the compression scheme optimization analyzed in Section~\ref{sec: bit rate allocation}. After the R-wave detection, we compress ECG signal according to the optimization results obtained from the proposed method. As a comparison, the compression without optimization is also performed. The mean square error between the original ECG and its reconstruction is taken as the metric. MSE with respect to the average bit rate per sample is plotted in Fig.~\ref{fig: bit rate opt effect}. 

From Fig.~\ref{fig: bit rate opt effect}, we can observe that the more bits enable us compress ECG signal at a smaller MSE. Furthermore, the optimum scheme can successfully reduce the average bit consumption by 1bit.

\begin{figure}[h]
\centering
\includegraphics[width=1\linewidth] {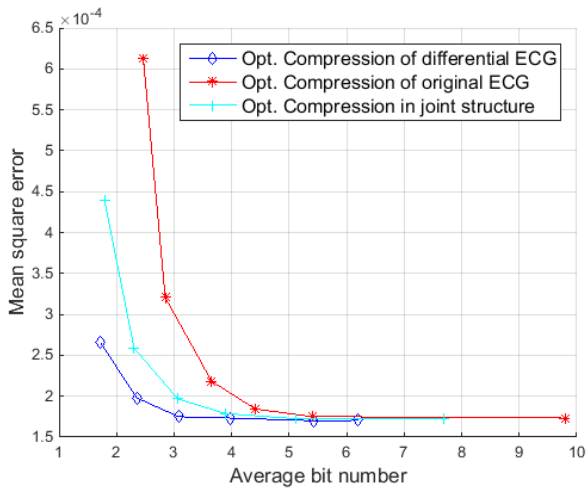}
\setlength{\abovecaptionskip}{3pt plus 3pt minus 2pt}
\caption{Comparison in bit rate allocation optimization for differential, original and joint ECG compression structures. }
\label{fig: diff orig joint}
\end{figure}


In the experiments, we compare the performance of the three compression structures proposed in Section~\ref{sec: ecg compression}. {MSE-bit rates} curves are taken as the metric in the performance comparison. A point in a {MSE-bit rates} curve informs us the minimum MSE achieved at a given number of bits. The simulation results from the three structures are plotted in Fig.~\ref{fig: diff orig joint}. 

From Fig.~\ref{fig: diff orig joint}, the differential compression structure consumes the smallest number of bit rate on average. The structure for compressing original ECG signal needs the largest number of bits among the three. Bit consumption by the joint scheme is in the middle. From the system block diagrams shown in Fig.~\ref{fig: original ecg scheme}, Fig.~\ref{fig: diff ecg scheme} and Fig.~\ref{fig: joint ecg scheme}, we also realize that differential compression structure is built with more complexity; the joint structure requires less complexity; and the original ECG compression structure can be implemented with the smallest complexity. 

\begin{figure}[h]
\centering
\includegraphics[width=1\linewidth] {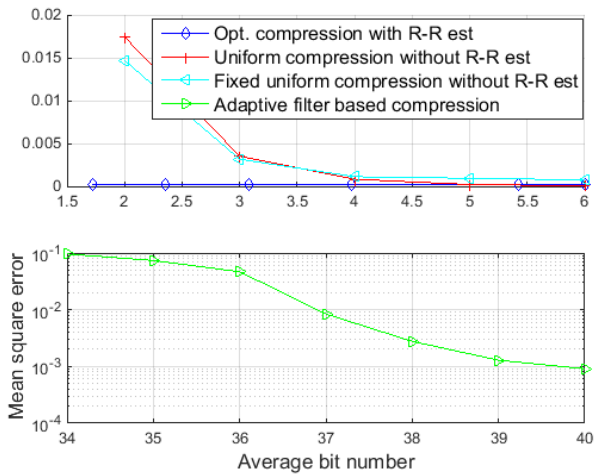}
\setlength{\abovecaptionskip}{3pt plus 3pt minus 2pt}
\caption{Comparison in bit rate allocation optimization with the methods in literature.}
\label{fig: comparison with literature}
\end{figure}

Furthermore, we compare our ECG compression algorithm with the ones in literature. Uniform quantization is a basic method of ECG compression. We consider two cases of uniform quantizations. First, the dynamical range of ECG signal is calculated in each segment. Second, the dynamical range is calculated only in the first segment. Very recently, adaptive filtering technology is employed in compressing ECG signal which is also investigated in the compression. Fig.~\ref{fig: comparison with literature} presents the performance comparison. 



From Fig.~\ref{fig: comparison with literature}, our ECG compression scheme achieves the smallest bit rate. Uniform quantization with dynamical range updating generates the larger MSE than our method. The uniform quantization without updating the dynamical range generates even larger error than the one with dynamical range updating. The adaptive filter based compression is able to achieve small error while the cost of bit rate is quite high. The high bit rate cost is the result of adaptive filter coefficients transmission.  

\section{Conclusion}
\label{sec:conclusion}



In this paper, we investigate bit rate optimization for remote ECG monitoring systems. ECG signal does not have white spectrum which means ECG samples are not independent to each other. Generally speaking, vector quantization has higher efficiency in compressing the signal with non-white spectrum. We consider the framework combining Huffman coding and DCT. The traditional Huffman plus DCT compression scheme use fixed parameters, such as quantization bits, quantization zones, and fixed bit rates. Different from the traditional scheme, we adaptively change the parameters for minimizing compression loss under a constraint of bit rate.

In the vector quantization, we need to partition ECG signal into segments. We partition ECG signal based on R-to-R periods. The segmentation generates a more stable statistics between ECG segments. The stability in statistical feature means more consistent optimum codebooks. Due to the high consistency, the computation used in optimizing codebook is significantly reduced. Following the analysis in ECG signal partitioning, we minimize the ECG reconstruction error under a constraint of a bit rate. These results obtained in this paper is beneficial for reducing the expenditure of bit rate which is important to build mobile ECG monitoring systems.

{\vspace{-0mm}
\scriptsize
\bibliography{chann_emu_InfoCom2013}
}
\bibliographystyle{IEEEtran}

\end{document}